\documentclass[preprint,12pt]{elsarticle}
\usepackage{graphics,bm}
\usepackage{amssymb}
\usepackage{epsfig}
\usepackage{epsf}
\usepackage{graphicx}
\usepackage{mathrsfs}
\usepackage{amsfonts}
\makeatletter

\newcommand{\beqa}{\begin{eqnarray}}
\newcommand{\eeqa}{\end{eqnarray}}
\newcommand{\beq}{\begin{equation}}
\newcommand{\eeq}{\end{equation}}

\journal{Journal of Magnetism and Magnetic Materials}
  
\begin{document}
\begin{frontmatter}

\title{Dynamical quenching of tunneling in molecular magnets}
\author[mjs]{Mar\'ia Jos\'e Santander}
\ead{maria.jose.noemi@gmail.com}
\author[focal]{Alvaro S. Nunez}
\ead{alnunez@dfi.uchile.cl}
\author[arm]{A. Rold\'an-Molina}
\author[rvt1]{Roberto E. Troncoso}
\ead{r.troncoso.c@gmail.com}
\address[mjs]{Recursos Educativos Qu\'antica, Santiago, Chile; Departamento de F\'isica, Universidad de Santiago de Chile and CEDENNA,
Avda. Ecuador 3493, Santiago, Chile}
\address[focal]{Departamento de F\'isica, Facultad de Ciencias F\'isicas y 
Matem\'aticas, Universidad de Chile, Casilla 487-3, Santiago, Chile}
\address[arm]{Instituto de F\'isica, Pontificia Universidad Cat\'olica de Valpara\'iso, Avenida Universidad 330, Curauma, Valpara\'iso, Chile}
\address[rvt1]{Centro para el Desarrollo de la Nanociencia y la Nanotecnolog\'ia, CEDENNA, Avda. Ecuador 3493, Santiago 9170124, Chile; Departamento de F\'isica, Universidad T\'ecnica Federico Santa Mar\'ia, Avenida Espa\~na 1680, Valpara\'iso, Chile}

\begin{abstract}
It is shown that a single molecular magnet placed in a rapidly oscillating magnetic field displays the phenomenon of quenching of tunneling processes. The results open a way to manipulate the quantum states of molecular magnets by means of radiation in the terahertz range. Our analysis separates the time evolution into slow and fast components thereby obtaining an effective theory for the slow dynamics. This effective theory presents quenching of the tunnel effect. In particular, stands out its difference with the so-called coherent destruction of tunneling. We support our prediction with numerical evidence based on an exact solution of the Schr\"odinger's equation.
\end{abstract}

\begin{keyword}
Tunnel effect; Topological Quenching; Magnetization Dynamics
\end{keyword}

\end{frontmatter}

\section{Introduction}\label{sec: Introduction}

Since a few decades molecular magnetic materials have arisen as a new test ground for several phenomena in quantum behavior of finite size magnetic systems  \cite{sessoli1993,Sessoli2006,wernsdorfer1999, sessoli2003,Friedman}. Molecular magnets have attracted attention due to their potential in the implementation of several molecular spintronic devices\cite{rocha2005,Wernsdorfer2008,Sanvito}. The long spin coherence times displayed turn them into promising candidates in the context of quantum computing, where the molecule spin is used to encode q-bits\cite{Loss,Friedman,Coronado}. 
 
The quantum mechanical degrees of freedom associated with molecular magnets can be manipulated and controlled with great accuracy by the application of external magnetic fields\cite{Sessoli2006,wernsdorfer1999,sessoli2003,Friedman}. It has been proposed\cite{Loss} that suitable manipulation of molecular magnets with time dependent magnetic fields can be used to control the population of  their quantum states thereby paving the road toward an implementation of a quantum computation scheme known as the Grover's algorithm. Within the same framework in this work we propose a way to control the quantum mechanical state of a molecular magnet by means of a rapidly varying magnetic field. We will show that radiation within the range of terahertz frequencies can be used to quench the quantum state of a molecular magnet thereby providing a useful tool for potential applications. There is previous work relating radiation in the terahertz range with magnetic properties of materials, for example terahertz radiation has been used to control the spin waves of antiferromagnets\cite{Kampfrath2011,Takayoshi} and in Ref. \cite{Kampfrath2013} hybrid magnetic structures have been used to generate radiation in the terahertz range. Our predictions allow an extension of such control into the subject of molecular magnets. We study the dynamics of the magnetization of a molecular magnet exposed to circularly polarized terahertz radiation. In response to such perturbations the system can display a quenching of the tunneling rate between energetically equivalent states giving rise to a trapping of the quantum mechanical state. This effect is analogous to an interesting effect in classical mechanics the trapping of a classical particle that can be achieved by introducing a rapidly oscillating potential $V(x,t)=V_0(x)+V_1(x,t)$ \cite{landau,kapitza}. Under the action of such forces the slow dynamics of the particle is trapped by an effective potential $V_{\rm eff}=V_0(x)+\overline{F^2}/(2m \omega^2)$ where $F$ is the force associated with the oscillating potential, $F=-\partial_x V_1(x,t)$,  $\omega$ its  frequency and the bar represents an average over an entire cycle of the oscillating force. Based on those ideas it was proposed and demonstrated by Kapitza\cite{kapitza} that a pendulum with a rapidly vibrating point of suspension would be stabilized in the upward position. Once stabilized the pendulum was shown to display small oscillations around its new equilibrium configuration. The main result of this paper is that a similar result holds for the quantum mechanical state associated with the spin of a molecular magnet. In this sense we can say that the prediction corresponds to a {\em Kapitza effect in Hilbert space}.  Instead of promoting transitions between states, the high frequency radiation traps the state of the spin in a given configuration keeping it from describing  tunnel transitions into other configurations. A quantitative statement of this effect is encoded in the tunneling time (the time that it takes for the system to tunnel from one minima to another) which is seen to diverge in certain circumstances. In response to the oscillatory disturbance the tunnel effect is suppressed and the states are frozen in a given configuration. A similar effect of suppression of tunneling has been reported in the literature concerning spins where the effect is attributed to interference of Berry's phases of different paths associated with tunneling. In this context the effect has been dubbed quenching of the tunnel amplitude\cite{garg1993,wernsdorfer1999}.

\section{Effective slow dynamics}
As is common in the theoretical studies of molecular magnets our study is based on the reduction of the electronic degrees of freedom, to some effective low-energy Hamiltonian, based entirely on localized spin degrees of freedom of the magnetic ions within the molecule\cite{Auerbach1994}.
%
In this context, the total energy has two contributions $\mathcal{H}=\mathcal{H}_0+\mathcal{H}_1$, one arising from the intrinsic anisotropy\cite{Sessoli2006}, in the form:
\begin{equation}
\mathcal{H}_0=-\mathcal{D} S^2_x+\mathcal{E}(S^2_z-S^2_y).
\end{equation} 
For $\mathcal{D}>\mathcal{E}$ this Hamiltonian represents a quantum spin with easy axis along the $x$ direction and a hard axis along the $z$ direction. The case ${\cal E}=0$ has been previously studied in Ref. \cite{Hemmen}. This Hamiltonian can be used as a model for describing the magnetic degrees of freedom of several single molecule nanomagnets\cite{Sessoli2006}. Among them the most widely used ones are the Mn$_{12}-ac$ molecule\cite{Friedman} (with $S=10$, $\mathcal{D}=0.55\; K$ and $\mathcal{E}=0.02\; K$), the molecular complex Fe$_8$ molecular magnet\cite{wernsdorfer1999} (with $S=10$, $\mathcal{D}=0.29\;K$ and $\mathcal{E}=0.05\;K$)  and Ni$_4$\cite{delBarco2004} (with $S=4$, $\mathcal{D}=0.75\;K$ and vanishingly small $\mathcal{E}$).
The spin is perturbed by a circularly polarized  time dependent  external field\cite{Takayoshi}:
\begin{equation}
\mathcal{H}_1=-h(\cos\omega t \; S_x+\sin\omega t\;  S_y)
\end{equation}
Here $\omega$ is the frequency of the oscillation of the magnetic field in the $x,y$ plane and $h=g\mu_B H$ with $H$ being the amplitude of the oscillating magnetic field and $g$ the gyromagnetic ratio (of order 1 in the examples given). Regarding the geometry we can say that it is dominated by the anisotropy terms in the Hamiltonian. To apply our theory in the experimental setting the incident radiation must be polarized in the plane perpendicular to the hard axis. In a molecular magnet based crystal this can be achieved by selecting the crystal orientation with respect to the incident light. In the remaining parts of this article we will address the problem of how the spin responds to the perturbation in the limit of large $\omega$.
\begin{figure}[htbp] 
   \centering
   \includegraphics[width=4in]{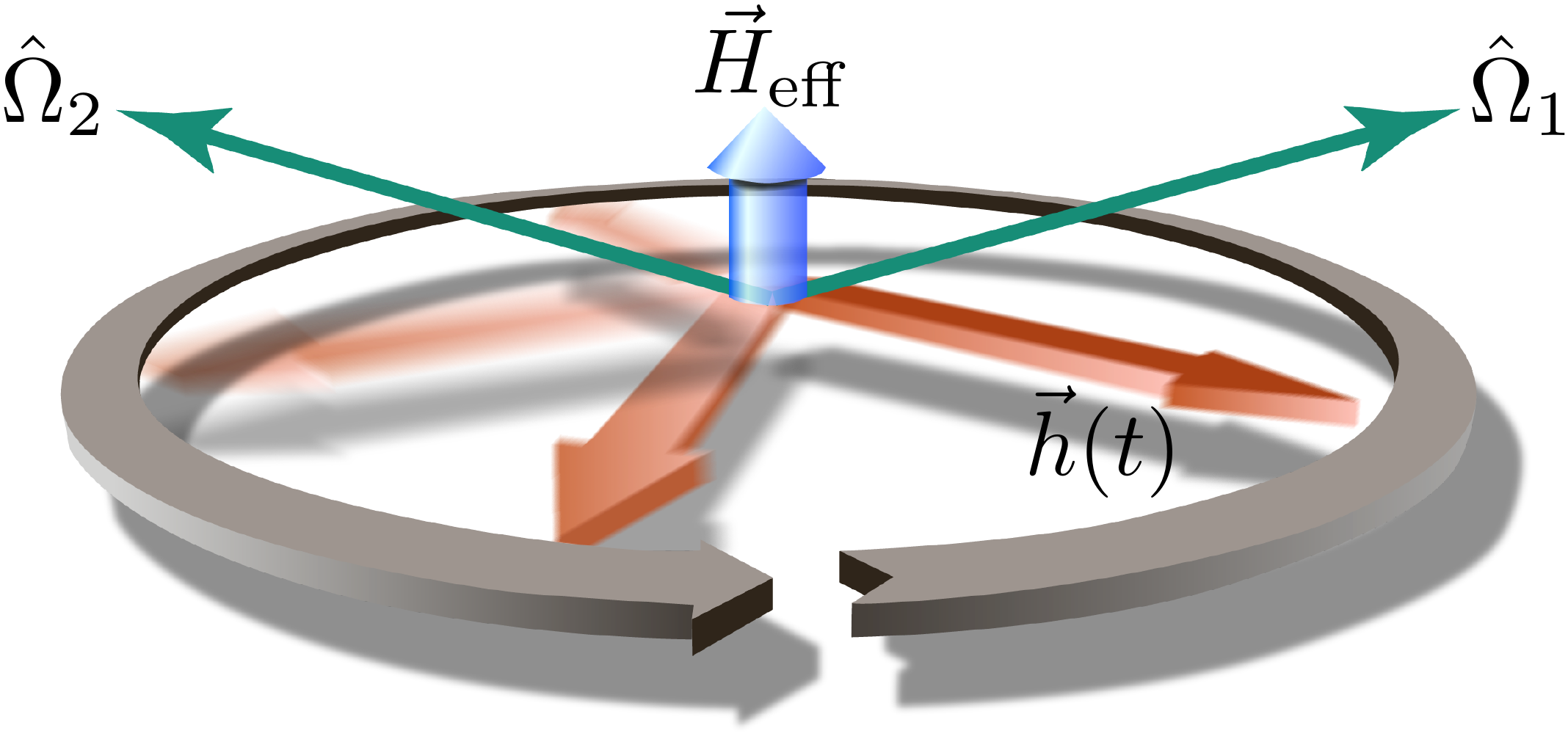} 
   \caption{Schematics of the proposed arrangement. A rapidly rotating magnetic field, $\vec{h}(t)$, in the plane induces a slow dynamics characterized  by a stationary effective field, $\vec{H}_{\rm eff}$. The direction of the effective magnetic field is perpendicular to the plane. The slow dynamics is characterized by two degenerate classical ground states $\hat{\Omega}_1$ and  $\hat{\Omega}_2$. In general, there are oscillations between those states mediated by quantum tunneling. However, for certain values of the parameters the quantum mechanical oscillations between those two states are quenched.}
   \label{fig:schematics}
\end{figure}
In the special case of spin $1/2$ all the anisotropic contributions reduce to the identity.  The resulting behavior has the characteristic form of Rabi oscillations\cite{Rabi}. For higher spin the interaction between the magnetic moment and a radiation field is  more complex and has been addressed in several references. For instance in Ref. \cite{Wernsdorfer2003} the effect of photon assisted tunneling events was reported in $\mathrm{Fe}_8$ samples irradiated with circularly polarized light. In Ref. \cite{delBarco2004} microwave spectroscopy was used to reveal quantum superpositions of high spin states ($S=4$) in $\mathrm{Ni}_4$. 

\begin{figure}[htbp] 
   \centering
   \includegraphics[width=4in]{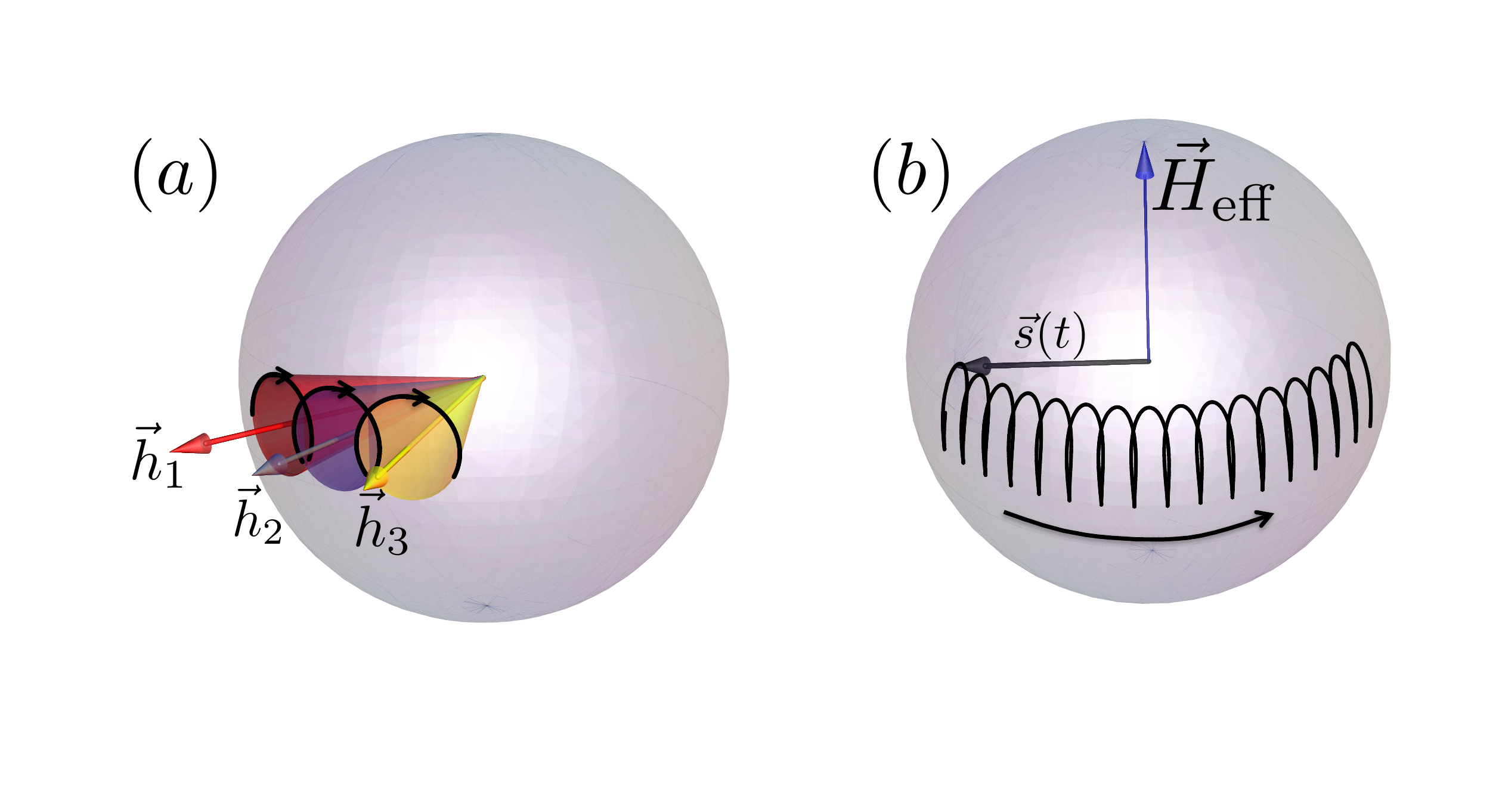} 
   \caption{Semiclassical origin of the effective magnetic field. (a) the spin precesses around an initial field $\vec{h}_1$. This precession is interrupted when the magnetic field changes into $\vec{h}_2$. Under the new field the precession axes is changed and a new precessional motion is described. A second change into $\vec{h}_3$ changes once again the precession axis. The net result of this sequence of changes is an effective precession around the $z$ axis. (b) Exact solution of the Landau-Lifshitz equation for $\omega=5 h$. The exact results match the qualitative argument of panel (a). The results are consistent with a precession around an effective magnetic field pointing in the $z$ axis.}
   \label{fig: precession}
\end{figure}

It is possible to derive a general treatment of a quantum system driven by rapidly varying potentials\cite{Rahav}. In general, the goal is to find an effective equation for the dynamics spanned by the 
Schr\"odinger equation:
\begin{equation}
i\hbar \partial_t|\Psi\rangle=\mathcal{H}|\Psi
\rangle
\label{eq: Schrodinger}
\end{equation}
As in the classical case\cite{Nayfeh}, the dynamics of the wave vector can be separated into two components, a fast one that varies in a period of the potential and a slow one that evolves at a slower pace. 
The method consists in an extension of the method of multiple scales from classical mechanics into quantum mechanics. We separate the slow and fast dynamics, by means of a unitary transformation, and proceed to write an effective theory that involves only the slow variables. The slow dynamics is affected by the rapid motion and is described by an effective Hamiltonian. We start from the time dependent Hamiltonian $\mathcal{H}$ and perform a time dependent unitary transformation $\exp(i\mathcal{F}(t))$. 
The basic idea is to absorb the time dependency of the Hamiltonian into the operator $\mathcal{F}(t)$ and to obtain an effective time independent Hamiltonian given by:
\begin{equation}
\mathcal{H}_{\rm eff}={\rm e}^{i\mathcal{F}}\mathcal{H}{\rm e}^{-i\mathcal{F}}+i\hbar\left(\frac{\partial {\rm e}^{i\mathcal{F}}}{\partial t}\right){\rm e}^{-i\mathcal{F}}.
\label{eq: eff hamil0}
\end{equation}
To find the specific representations of  $\mathcal{H}_{\rm eff}$ and $\mathcal{F}$ we write them as  power series in $1/\omega$: 
\begin{eqnarray}
\mathcal{H}_{\rm eff}&=&\mathcal{H}^{(0)}_{\rm eff}+\frac{1}{\hbar\omega}\mathcal{H}^{(1)}_{\rm eff}+\frac{1}{(\hbar\omega)^2}\mathcal{H}^{(2)}_{\rm eff}+\cdots\\
\mathcal{F}&=&\frac{1}{\hbar\omega}\mathcal{F}^{(1)}+\frac{1}{(\hbar\omega)^2}\mathcal{F}^{(2)}+\cdots
\end{eqnarray}
Expanding  Eq. (\ref{eq: eff hamil0}) and equating both sides order by order we found recursive relations between $\mathcal{F}^{(i)}$  and $\mathcal{H}^{(i)}_{\rm eff}$. The expression for $\mathcal{F}^{(i)}$ is found by enforcing that every term in the expansion of the effective Hamiltonian be time independent. A lengthy, but straightforward calculation leads to the following first contributions in the expansion for $\mathcal{F}$:
\begin{eqnarray}
\mathcal{F}^{(1)}&=& -h\left(\sin\omega t S_x-\cos\omega t S_y\right)
\end{eqnarray}
and
\begin{eqnarray}
\mathcal{F}^{(2)}&=& -ih\left(\cos\omega t [S_x,\mathcal{H}_0]+\sin\omega t [S_y,\mathcal{H}_0]\right),
\end{eqnarray}
while for  the Hamiltonian we obtain:
\begin{eqnarray}
\mathcal{H}^{(1)}_{\rm eff}&=& -\frac{h^2}{2} S_z
\end{eqnarray}
and
\begin{eqnarray}
\mathcal{H}^{(2)}_{\rm eff}&=& \frac{h^2}{4}\left([[S_x,\mathcal{H}_0],S_x]+[[S_y,\mathcal{H}_0],S_y]\right)
\end{eqnarray}
Collecting the contributions up to second order we obtain an effective Hamiltonian:
\begin{equation}
\mathcal{H}_{\rm eff}=\mathcal{H}_0-\frac{h^2}{2\hbar\omega}S_z+\frac{h^2}{4\hbar^2\omega^2}\left(
\left[\left[S_x,\mathcal{H}_0\right],S_x\right]+\left[\left[S_y,\mathcal{H}_0\right],S_y\right]
\right),
\label{eq: eff hamil}
\end{equation}
In the limit of small $\mathcal{D}S/\hbar\omega$, i.e. for sufficiently rapid variations of the magnetic field, the effective Hamiltonian corresponds to the original stationary contribution to the energy plus an effective field pointing in the direction perpendicular to the plane of polarization of the magnetic field. For common molecular magnets such as the ones described above the condition $\mathcal{D}S/\hbar\omega\ll 1$ sets the stage in a magnetic field oscillating in the terahertz range\cite{Friedman}. The direction and magnitude of this effective field is independent of $\mathcal{H}_0$. We can explain the origin of this contribution in terms of the following semi classical argument. The equation of motion of a spin in absence of any anisotropy is simply given by the Landau-Lifshitz equation, 
$\frac{d\vec{S}}{dt}=\vec{S}\times \vec{h}(t)$,
where $\vec{h}(t)$ correspond to the oscillating magnetic field. The motion can be understood as a succession of changes each of one is aprecession around the moving field. As shown in Fig. \ref{fig: precession} the complicated sequence of small precessions leads to a simple motion, a steady precession around the $z$ axis. This can be associated with an effective field pointing along the same axis. As shown in Eq. (\ref{eq: eff hamil}) this field is generally present regardless of the base Hamiltonian.
This is not true for the following terms in the expansion. The second order term depends explicitly upon the stationary Hamiltonian. A fundamental result of this paper follows after a direct calculation of this second order term, it can be easily verified that the contribution to order $\omega^{-2}$ merely shifts the values of the constants $\mathcal{D}$ and $\mathcal{E}$ with a correction of order $h^2/(\hbar\omega)^{2}$. Hereafter we will denote those corrected values by $\tilde{\mathcal{D}}$ and $\tilde{\mathcal{E}}$ respectively. We conclude that the effective theory describes a spin with corrected anisotropy energies exposed to an effective magnetic field in the direction perpendicular to the plane of polarization of the time varying magnetic field. From this result we can infer several properties of the slow dynamics described by the quantum states. 

For spin $1/2$ we can use our result in Eq. (\ref{eq: eff hamil}) in a direct way. Since the Hamiltonian $\mathcal{H}_0$ reduces to the identity (a property of the spin $1/2$ operators) the evaluation of the effective Hamiltonian is straightforward. We have a very simple result, namely:
\begin{equation}
\mathcal{H}_{\rm eff}=-\frac{h^2}{2\hbar\omega}S_z,
\label{eq: eff hamil}
\end{equation}
which provides a two level system with energy levels equal to $\pm\frac{h^2}{4\hbar\omega}$. A system prepared in a mixed state will oscillate\cite{Rabi} between the eigenstates with a frequency determined by this splitting. For greater spins the anisotropy contribution plays an essential role in the dynamics as we will show in the next section.

\section{Dynamical suppression of quantum tunneling}\label{sec: quenching}
The slow dynamics reduces to the well-known problem of a spin under the action of an external magnetic field, $H_{\rm eff}$, along the hard axis,  whose strength is given by $g\mu_BH_{\rm eff}=h^2/(2\hbar\omega)$. In this way we see that the resulting Hamiltonian is well studied in several contexts\cite{Sessoli2006}.  We can readily infer a number of properties regarding the behavior of the quantum moment by exploiting this analogy. 
We start by analyzing the  classical limit. It is clear that the ground state is doubly degenerate. Using spherical coordinates, $\hat{\Omega}=(\sin\theta\cos\phi,\sin\theta\sin\phi,\cos\theta)$, as shown in Fig. \ref{fig:schematics},  we find the two minima located at:
\begin{equation}
\cos\theta_0=\frac{h^2}{4 S \hbar \omega (\tilde{\mathcal{D}} +\tilde{\mathcal{E}})}
\end{equation} 
and $\phi_{1,2}=0,\pi$. The degeneracy of the classical ground states is lifted by quantum fluctuations associated with the tunnel effect between the two minima. The resulting energy splitting $\Delta$ between the two lowest lying states characterizes the tunneling time by the relation $T=\hbar/\Delta$. A detailed calculation of this splitting has been made and the results are highlighted in Fig. 4. A state initially prepared around one of the minima will oscillate, tunneling across the energy barrier, into the other minima in a characteristic time $T$. 
It is possible to find the tunneling time by direct diagonalization of the effective Hamiltonian. The tunnel splitting $\Delta$ oscillates as a function of the effective magnetic field. For certain specific values of the effective field a phenomenon known as the suppression or quenching of the tunnel effect can be observed. For those values of $H_{\rm eff}$ the tunnel splitting vanishes and the tunneling time diverges. The state is, therefore, trapped in a given state. The quenching of the tunneling processes, revealed by the reduction of the energy splitting between the two lowest lying energy states, has an origin that can be traced back to the interference between different tunneling paths. Such behavior is better understood in terms of a semiclassical analysis of the tunneling process as is given by the instanton technique\cite{garg1993,Chen 2002,instantones1,instantones2}.
The origin of the oscillation of the tunneling gap is the interference between the Berry's phases associated with complementary paths that accomplish the reversal. The result is\cite{garg1993} $\Delta=\Delta_0\cos(S\Omega)$ where $\Omega$ is the solid angle subtended by the complementary paths and  $\Delta_0$ is a monotonous function of $h$. Whenever $S\Omega=(2n+1)\pi/2$ for integer $n$ the splitting vanishes. It is important to note that the dynamical supression of the Rabi oscillations due to the rapidly oscillating magnetic field arises due to the interference between tunneling amplitudes associated with different reversal  paths and is, therefore, different in origin from the well-known coherent suppression of tunneling that has been predicted\cite{CST} in one dimensional systems. 
This expression makes explicit the oscillations of the gap as function of the external driving field. However, the most important fact is that  it emphasizes their origin in the Berry's phase interference of complementary paths that achieve the tunnel reversal. For this reason, the effect is known as Berry's phase interference quenching of the tunnel effect\cite{garg1993}. 
This semiclassical argument ought to be contrasted with the exact diagonalization of the Hamiltonian given in Eq. (\ref{eq: eff hamil}) for different values of the magnetic field intensity, $h$. After a direct numerical diagonalization, it is possible to calculate the energy gap between the two lowest lying states. This gap is shown, for $S=1$, $S=3/2$, $S=2$ and $S=5/2$, in the continuous lines in the Fig. \ref{fig:tunneling time}  where the tunneling times are plot as functions of $h^2/(2\hbar\omega)$. In those figures it is evident the oscillatory behavior displayed by the gap and the quenching associated with its zeroes (indicated by arrows).
 We conclude that the effective slow dynamics of a spin under a circularly polarized magnetic field will be described by oscillations between two states $|\Omega_1\rangle$ and $|\Omega_2\rangle$, where $\Omega_{1,2}$ correspond to the unit vector with polar angles $(\theta_0,\phi_{1,2})$. Here, the state $|\Omega\rangle$ corresponds to the spin coherent state oriented along the direction of $\Omega$\cite{Auerbach1994}.  Those oscillations correspond to tunneling events.  Furthermore, it can be concluded that those oscillations are quenched for certain values of the amplitude of the oscillating magnetic field. In the next section this prediction will be contrasted with exact numerical results. 

\section{Comparison with exact results}
\begin{figure}[htbp] 
   \centering
   \includegraphics[width=4in]{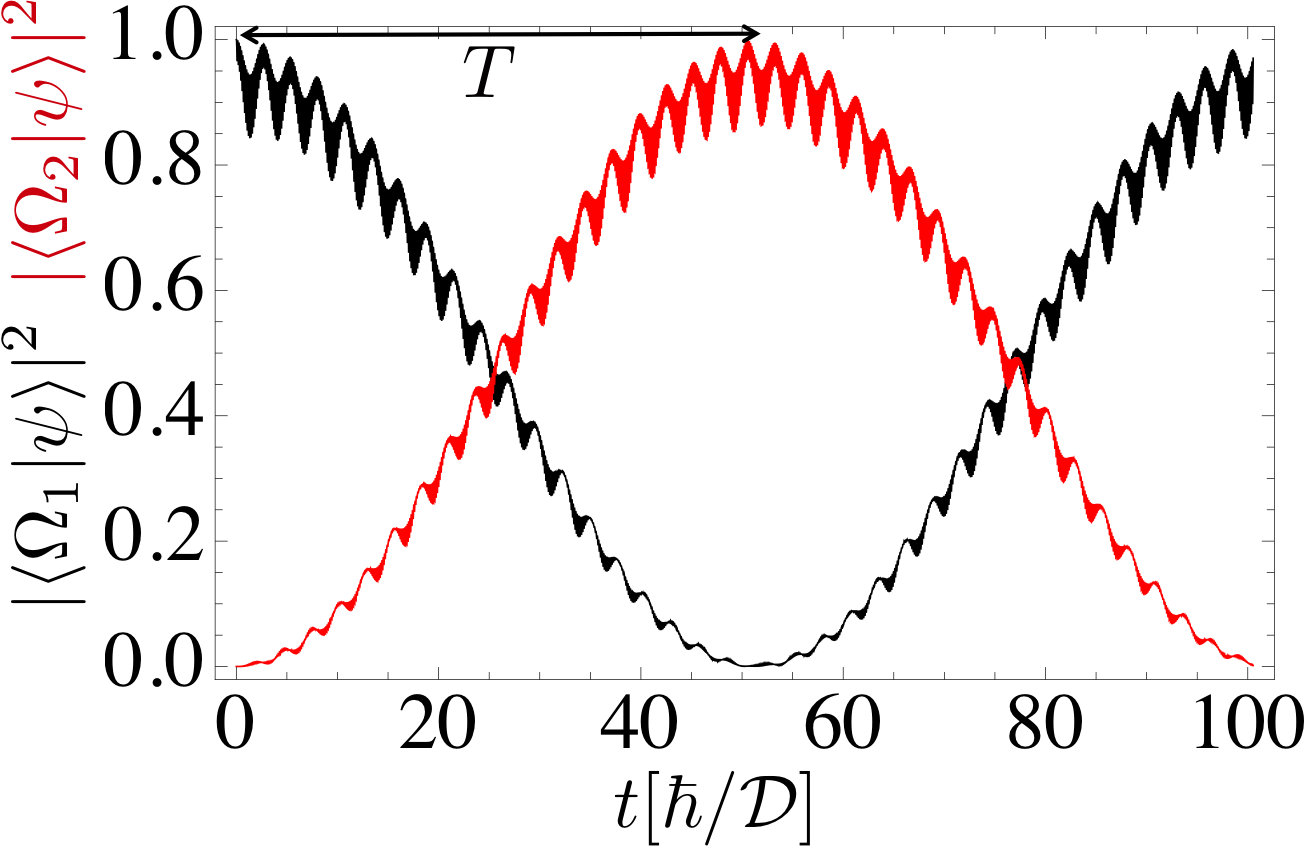} 
   \caption{Characteristic behavior of the projections $|\langle\Omega_1|\Psi\rangle|^2$ and $|\langle\Omega_2|\Psi\rangle|^2$ as functions of time. The initial state is $|\Omega_1\rangle$ and in the slow dynamics regime the system manifest well defined tunneling oscillations between the two classical ground states. The tunneling time, defined as the characteristic time that takes the system from one configuration to the other, is denoted by $T$. The slow dynamics is dressed by very fast oscillations.}
   \label{fig:ket1ket2}
\end{figure}

\begin{figure}[h!] 
\includegraphics[width=2.6in]{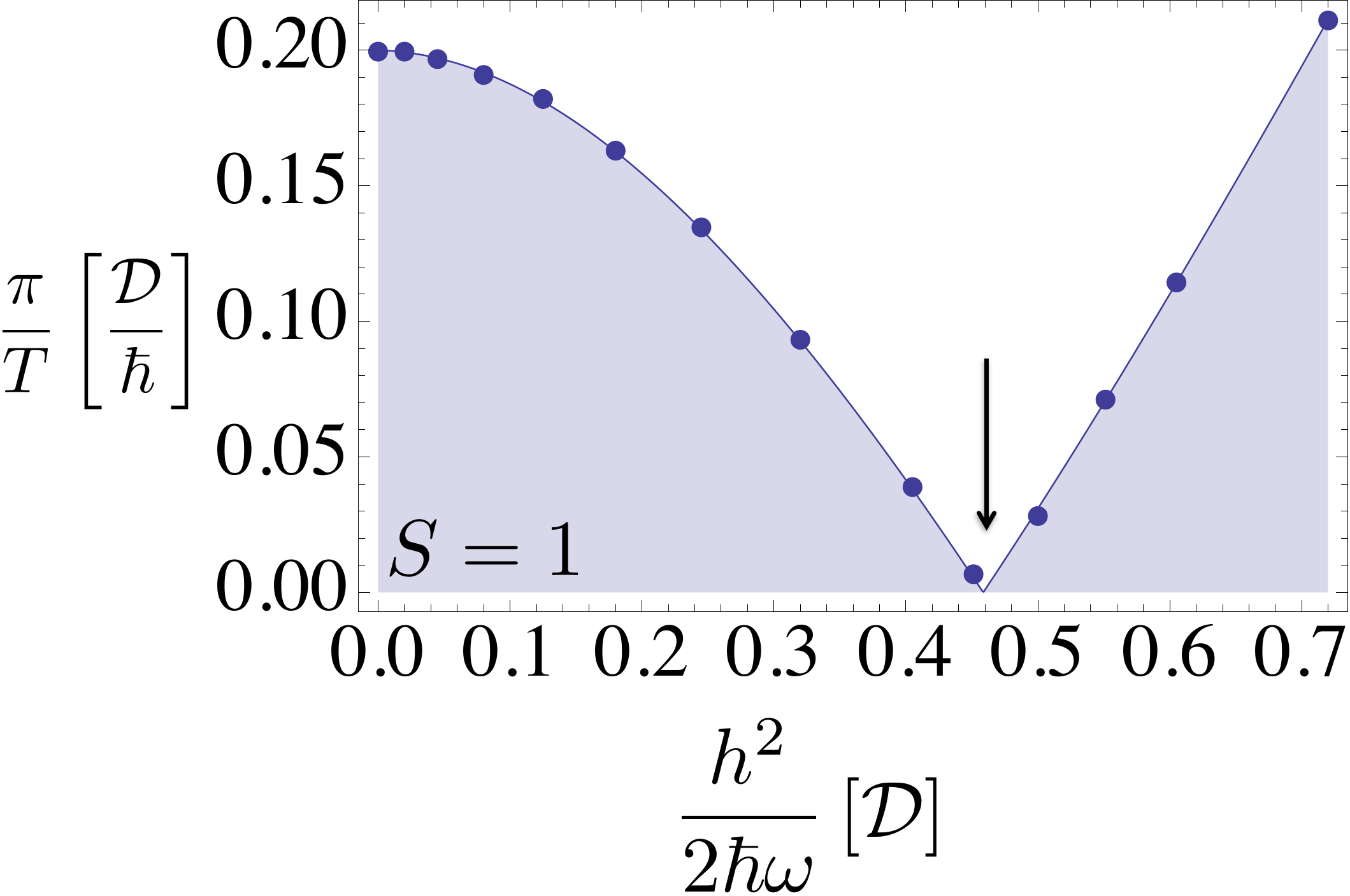}
\includegraphics[width=2.6in]{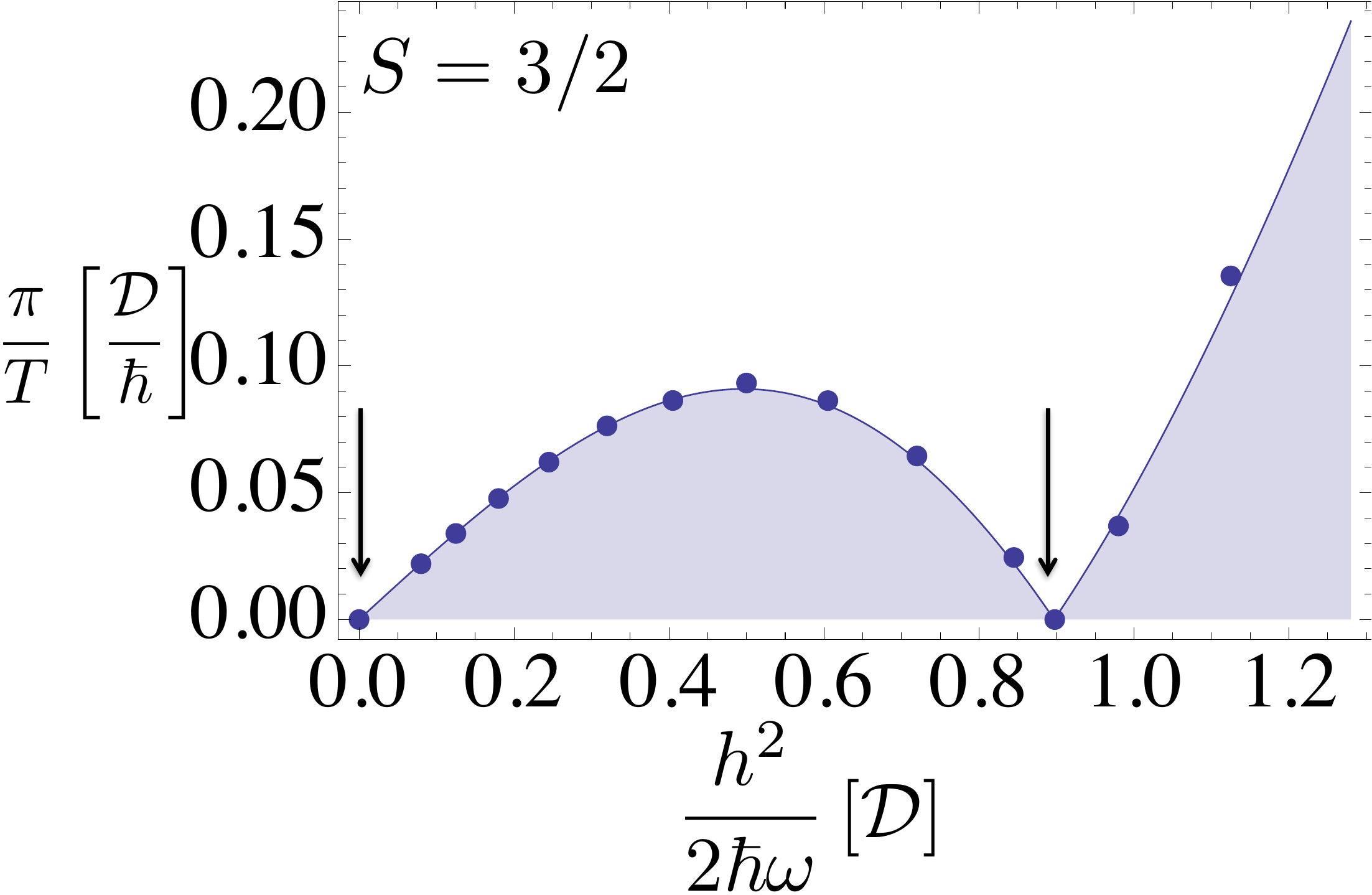}\\
\includegraphics[width=2.6in]{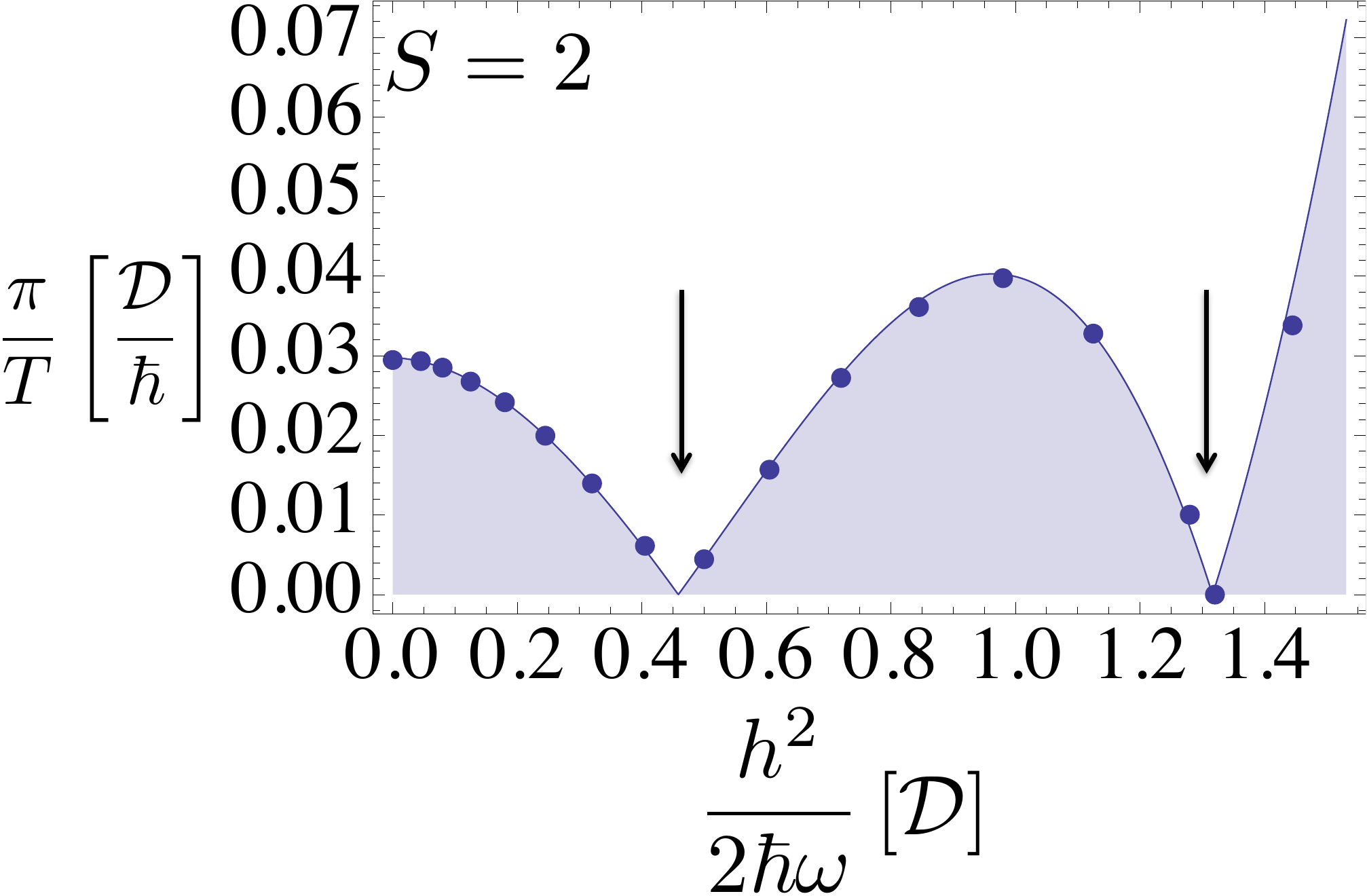}
\includegraphics[width=2.6in]{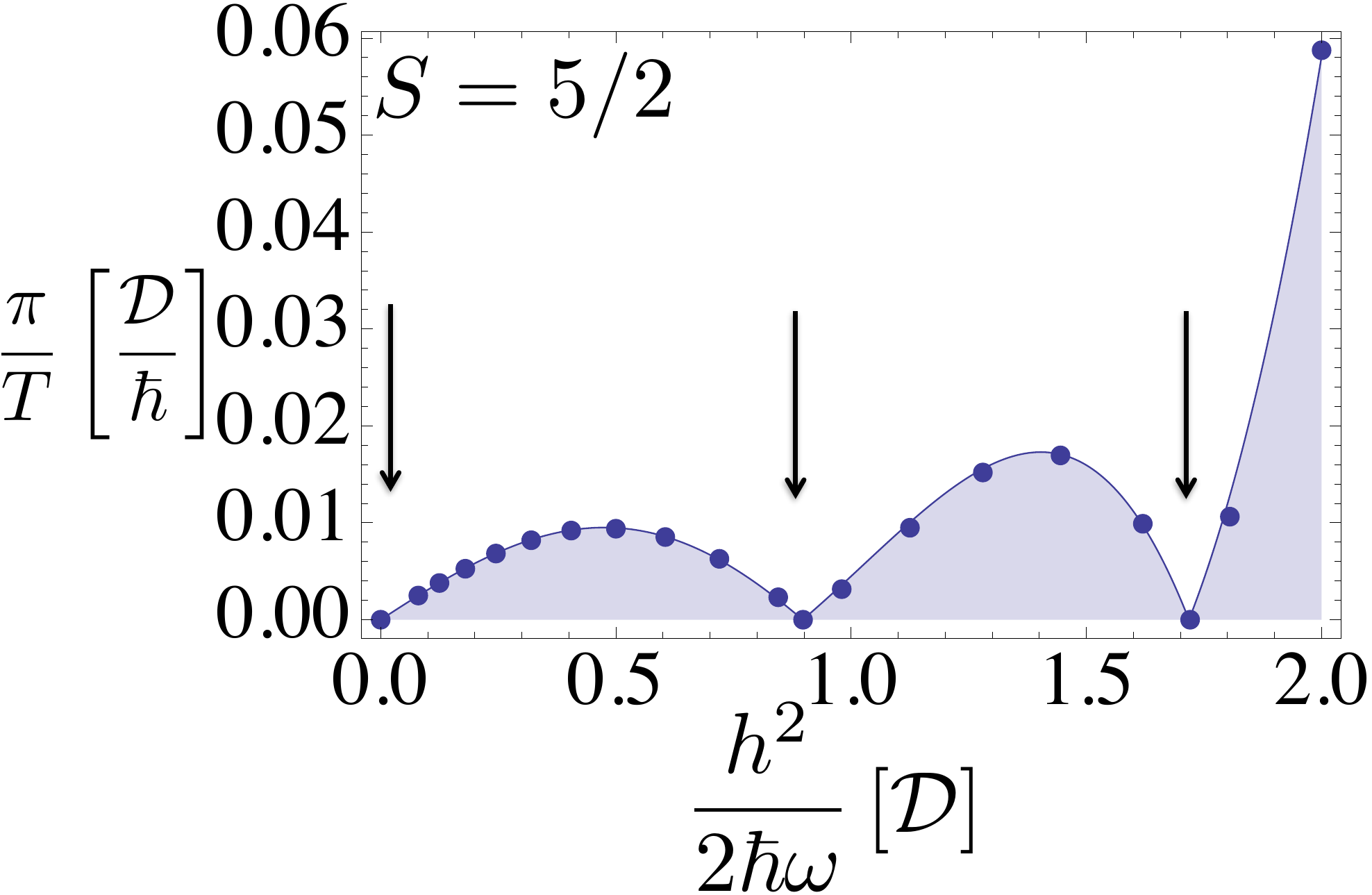}
   \caption{Tunneling time is shown for spin $S=1$, $S=3/2$, $S=2$, and $S=5/2$ as a function of the intensity of the magnetic field $h_{eff}=h^2/(2\hbar\omega)$. The tunneling time, obtained directly from the numerical solutions of Eq. (\ref{eq: Schrodinger}), is represented by the dots. It is compared with the predictions from the effective theory given by Eq. (\ref{eq: eff hamil}).  The effect of quenching of the tunnel effect implies a divergence of the tunneling time as indicated by the arrows. }
   \label{fig:tunneling time}
\end{figure}

We now contrast the predictions made so far with  exact numerical results. To that end we solve numerically the time dependent Schr\"odinger equation Eq. (\ref{eq: Schrodinger}).
 To characterize the slow dynamics generated by the Schr\"odinger equation we have prepared the state initially in the ket $|\Omega_1\rangle$ and let it evolve. We have computed the projection of the state vector $|\Psi(t)\rangle$ on the kets $|\Omega_1\rangle$ and $|\Omega_2\rangle$ with typical behavior shown in Fig. \ref{fig:ket1ket2} where we plot $|\langle \Omega_1|\Psi\rangle|^2$ and $|\langle \Omega_2|\Psi\rangle|^2$. We see that  the projection on the initial state $|\Omega_1\rangle$ starts at its maximum value, one, and then it is reduced to zero. 
Meanwhile the projection on the second state describes the opposite behavior. We remark that the oscillation in the amplitude is accompanied by a rapidly oscillatory component that corresponds to the fast dynamics. This behavior, reminiscent to that of a two level system, corresponds to the quantum tunneling between the states described by the states  $|\Omega_1\rangle$ and $|\Omega_2\rangle$. The characteristic time $T$ that takes a transition from one state to the other correspond to a direct quantification of the tunneling time. This time can be inferred directly from the numerical solution. For different values of the field we calculate such tunneling time. The results, for $S=1$, $S=3/2$, $S=2$ and $S=5/2$, are displayed in Fig. \ref{fig:tunneling time}. 

In the different calculations we have chosen units in which $\mathcal{D}=1$ and selected a  value for $\mathcal{E}=0.1\mathcal{D}$. For the numerical calculations we have selected the driving frequency as $\omega=50 \mathcal{D}/\hbar$. The results are in evident agreement  with those obtained from the effective theory. This provides strong evidence in favor of the validity of the effective Hamiltonian (Eq. (\ref{eq: eff hamil})) to describe an important aspect of the complicated dynamics displayed by the system. These exact results show that the tunnel effect is effectively quenched by the time dependent magnetic field for certain values of the field intensity and frequency. This quenching is made manifest by a divergence of the tunneling time associated with quantum transitions between the two degenerate classical ground states.

\section{Discussion}
In this paper we have presented a detailed study of the dynamics described by a quantum spin when exposed to a rapidly varying magnetic field. The problem is analogous to the pendulum, proposed by Kapitza, with a rapidly oscillating suspension point. Just like in the Kapitza's pendulum the motion of the quantum states is decomposed into two components. We have a rapid behavior whose characteristic time is given by the external potential and a slow contribution.  The slow dynamics is affected by the rapid motion and is described by an effective Hamiltonian.
 By performing this separation in a systematic fashion we have proved that the slow dynamics is accounted for by an effective Hamiltonian that has a Zeeman-like contribution in the direction perpendicular to the plane of polarization of the magnetic field. The effective Hamiltonian displays the interesting effect of  geometric-phase induced quenching of the tunnel effect that can be used to freeze the quantum states in well defined configurations. Just like the Kapitza's pendulum is trapped in the vertical position, unstable when the driving force is absent, the spin of the molecular magnet is trapped in a given quantum state. In this sense we  talk about a Kapitza effect in Hilbert space. This effect provides a tool to control the quantum states by means of high frequency fields. For a typical molecular nanomagnet our analysis sets the desired frequencies in the terahertz range. 
In addition we have presented numerical evidence, based on a direct solution of the Schr\"odinger equation, that confirms the predictions of the effective description. 
\section*{Acknowledgment}

The authors
acknowledge support from Fondo Nacional de Desarrollo
Cient\'ifico y Tecnol\'ogico Grant No. 1150072; Basal Program Center for Development
of Nanoscience and Nanotechnology (CEDENNA); and Anillo de Ciencia y Tecnonolog\'ia ACT Grant
No. 1117;

\end{document}